\documentclass[apj]{emulateapj}
\newcommand {\Lya}    {Ly$\alpha$}   %  Lyalpha
\newcommand {\Lyb}    {Ly$\beta$}    %  Lybeta
\newcommand {\Lyg}    {Ly$\gamma$}

\newcommand {\HI}     {\ion{H}{1}}   %  HI
\newcommand {\OVI}    {\ion{O}{6}}   %  OVI
\newcommand {\OVII}   {\ion{O}{7}}
\newcommand {\OVIII}  {\ion{O}{8}}
\newcommand {\CIII}   {\ion{C}{3}}   %  CIII
\newcommand {\NV}     {\ion{N}{5}}
\newcommand {\CIV}    {\ion{C}{4}}
\newcommand {\SiIV}   {\ion{Si}{4}}
\newcommand {\SiIII}  {\ion{Si}{3}}
\newcommand {\SiII}   {\ion{Si}{2}}

\newcommand {\CII}    {\ion{C}{2}}
\newcommand {\CI}    {\ion{C}{1}}

\newcommand {\kms}    {km~s$^{-1}$}
\newcommand {\NHI}    {$N_{\rm HI}$}

\newcommand {\NCIII}  {$N_{\rm CIII}$}

\newcommand {\NSiIII} {$N_{\rm SiIII}$}

\newcommand {\NNV}    {$N_{\rm NV}$}

\newcommand {\lam}    {$\lambda$}
\newcommand {\FUSE}  {{\it FUSE}}

\newcommand {\dndz}  {$d{\cal N}/dz$}

\newcommand {\etal}  {et~al.}

\slugcomment{ApJ submitted May 11, 2010}
\shorttitle{COS Observations of 1ES\,1553$+$113}
\shortauthors{Danforth \etal}

\begin{document}

\title{Hubble/COS Observations of the Ly\,$\alpha$ Forest toward the BL\,Lac Object 1ES\,1553$+$113\footnote{Based on observations made with the NASA/ESA {\it Hubble Space Telescope}, obtained from the data archive at the Space Telescope Science Institute. STScI is operated by the Association of Universities for Research in Astronomy, Inc. under NASA contract NAS 5-26555.}}
\author{Charles W. Danforth, Brian A. Keeney, John T. Stocke, J. Michael Shull, \& Yangsen Yao}
\affil{CASA, Department of Astrophysical \& Planetary Sciences, University of Colorado, 389-UCB, Boulder, CO 80309; danforth@casa.colorado.edu}

% ****************************************************************** %

\begin{abstract}
We present new moderate-resolution, far-ultraviolet spectra from the {\it Hubble Space Telescope}/Cosmic Origins Spectrograph (HST/COS) of the BL\,Lac object 1ES\,1553$+$113 covering the wavelength range $\rm 1135~\AA<\lambda<1795~\AA$.  The data show a smooth continuum with a wealth of narrow ($b<100$ \kms) absorption features arising in the interstellar medium (ISM) and intergalactic medium (IGM).  These features include 41 \Lya\ absorbers at $0<z_{\rm abs}<0.43$, fourteen of which are detected in multiple Lyman lines and six of which show absorption in one or more metal lines.  We analyze a metal-rich triplet ($\Delta cz\sim1000$ \kms) of \Lya\ absorbers at $z_{\rm abs}\approx0.188$ in which \OVI, \NV, and \CIII\ absorption is detected.  Silicon ions (\SiIII, \SiIV) are not detected to fairly strong upper limits, and we use the measured \SiIII/\CIII\ upper limit to derive an abundance limit $\rm(C/Si)\ge4\,(C/Si)_\sun$ for the strongest component of the absorber complex.  Galaxy redshift surveys show a number of massive galaxies at approximately the same redshift as this absorption complex, suggesting that it arises in a large-scale galaxy filament.   As one of the brightest extragalactic X-ray and $\gamma$-ray sources, 1ES\,1553$+$113 is of great interest to the high-energy astrophysics community.  With no intrinsic emission or absorption features, 1ES\,1553$+$113 has no direct redshift determination.  We use intervening \Lya\ absorbers to place a direct limit on the redshift: $z_{\rm em}>0.395$ based on a confirmed \Lya$+$\OVI\ absorber and $z_{\rm em}>0.433$ based on a single-line detection of \Lya.  The current COS data are only sensitive to \Lya\ absorbers at $z<0.47$, but we present statistical arguments that $z_{\rm em}\la0.58$ (at a $1\sigma$ confidence limit) based on the non-detection of any \Lyb\ absorbers at $z>0.4$.

\end{abstract}


%%%%%%%%%%%%%%%%%%%%%%%%%%%%%%%%%%%%%%%%%%%%%%%%%%
%% Figure 1
\begin{figure*}[t]
  \epsscale{1.1}\plotone{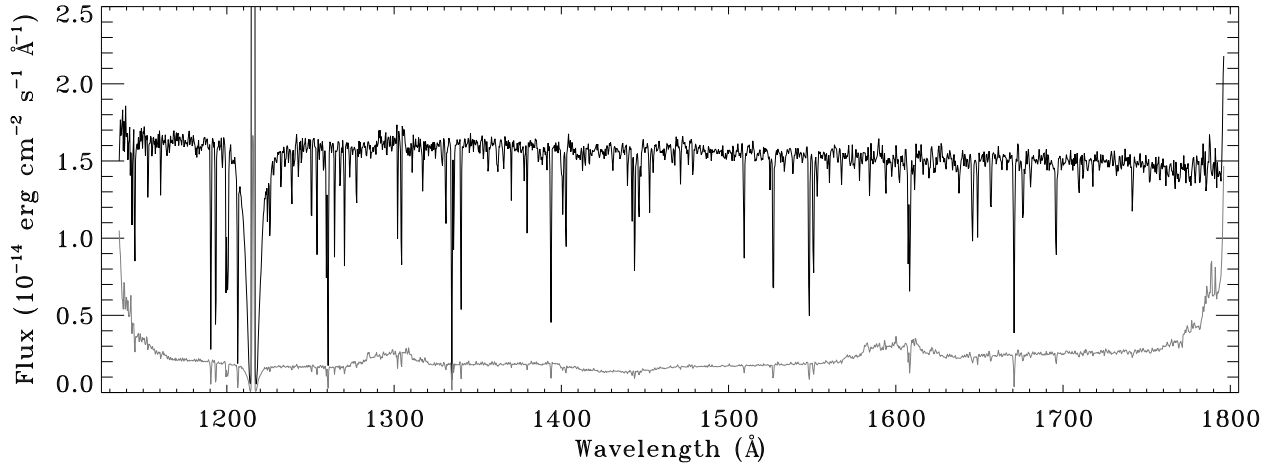} %%%%%%%%%%%%%
  \caption{Overview of the COS spectrum of 1ES\,1553$+$113.  The full
  COS/G130M$+$G160M dataset is shown, smoothed by 35 pixels ($\sim5$
  resolution elements).  Error is shown in gray.}
\end{figure*}

\section{Introduction}

The current interpretation of BL\,Lac objects \citep{Ghisellini85} is that they are active galactic nuclei (AGN) with a strongly relativistic jet pointed toward our line of sight.  As such, any line emission or accretion disk features seen in most other types of AGN could be masked by the bright jet if present in BL Lac objects.  Their spectra usually show a featureless power-law continuum extending from radio to X-ray wavelengths.  This spectral characteristic makes BL\,Lac objects ideal for observing intervening absorption features arising in the interstellar medium (ISM) and intergalactic medium (IGM).  Since their continuum is easily defined, they make excellent targets for studying weak metal-line systems and low-contrast, highly thermally broadened \HI\ absorbers \citep[e.g.,][]{Richter04,Lehner07,Danforth10}.

The BL\,Lac object 1ES\,1553$+$113 shows the characteristic featureless power-law spectrum and is one of the brightest known sources of extragalactic high-energy radiation from X-rays up to VHE (TeV) photons \citep{CostamanteGhisellini02}.  However, the featureless spectrum makes it difficult to determine the redshift of the object and hence its luminosity.  Indirect methods have given a wide range of limits for the redshift of 1ES\,1553$+$113; the nondetection of a host galaxy gave limits from $z_{\rm em}>0.09$ to $z_{\rm em}>0.78$ \citep{HutchingsNeff92,Scarpa00,Urry00,Carangelo03,Sbarufatti06,Treves07}.  The shape of the $\gamma$-ray spectrum observed by the {\it Fermi Observatory} and ground-based VHE detectors (HESS, MAGIC) constrains the redshift to values from $z_{\rm em}<0.4$ to $z_{\rm em}<0.8$ \citep{Ahronian06,Albert07,MazinGoebel07,Abdo10} based on assumptions about the intrinsic spectral energy distribution (SED) and pair-production interactions with the cosmic infrared background.  The only direct redshift determination \citep[$z_{\rm em}=0.36$;][]{MillerGreen83} was based on a spurious feature in low-resolution UV spectra from the {\it International Ultraviolet Explorer} (IUE).  The detection was later retracted \citep{FalomoTreves90}, but the erroneous redshift value lives on.

1ES\,1553$+$113 is of interest as a bright background continuum source for detecting intergalactic absorption along the sight line.  Bright X-ray sources are especially valuable for potentially detecting the long-predicted \OVII\ and \OVIII\ tracers \citep{Bregman07} of intergalactic gas at $T=10^6-10^7$~K.  Even for a bright X-ray source, the required integration times would be very long.  However, a sufficiently long IGM pathlength provided by a bright high-$z$ target would make the required observing time investment more attractive. 

In this paper, we present the first medium-resolution far-UV spectroscopic observations of 1ES\,1553$+$113 including {\it Hubble Space Telescope}/Cosmic Origins Spectrograph \citep[HST/COS][]{Green10,Osterman10} observations ($\lambda=1135-1795$~\AA) as well as archival data at $905-1187$~\AA\ from the {\it Far Ultraviolet Spectroscopic Explorer} \citep[\FUSE;][]{Moos00,Sahnow00}.  We confirm the featureless power-law nature of the spectrum over this wavelength range.  Absorption is seen in 42 intervening systems including 41 \Lya\ absorbers and six metal-line systems.  The frequency of IGM absorbers is consistent with larger surveys using \FUSE\ and HST/STIS data (Danforth \& Shull 2005, 2008; hereafter DS08), and the systems are spread across the entire redshift range covered by the combined COS/\FUSE\ dataset ($z\la0.47$).

The observations and data reduction techniques are discussed in \S2, and we present a preliminary catalog of absorption lines in \S3.  Our conclusions are presented in \S4.

%%%%%%%%%%%%%%%%%%%%%%%%%%%%%%%%%%%%%%%%%%%%%%%%%%%%%%%%%%%%%%%%%%%%%%%%%%%%
\section{Observations and Data Analysis}

Far-UV observations of 1ES\,1553$+$113 were carried out 2009 September 22 by HST/COS as part of the COS Guaranteed Time Observations (PID 11528, PI Green).  Five exposures were made in each of the G130M ($1135 < \lambda < 1480$~\AA) and G160M ($1400 < \lambda < 1795$~\AA) medium-resolution gratings ($R\approx18,000$) totalling 3.1 and 3.8 ksec, respectively.  Four central wavelength settings at each grating dithered known instrumental features along the spectrum and provided continuous spectral coverage over $1135 < \lambda < 1795$~\AA\ \citep[see][]{Green10,Osterman10}.  After retrieval from the archive, all ten exposures were reduced locally using {\sc CalCOS v2.11f}.

%% coaddition discussion
Flat-fielding, alignment, and coaddition of the processed exposures were carried out using IDL routines developed by the COS GTO team specifically for COS FUV data\footnote{See {\tt http://casa.colorado.edu/$\sim$danforth/costools.html} for our coaddition and flat-fielding algorithm and additional discussion.}.  First, the data were corrected for the most egregious instrumental features.  While attempts at a true ``flat-fielding'' of COS data show promise, the technique is not yet robust enough to improve data of moderate S/N.  However, we are able to correct the narrow $\sim$15\%-opaque features arising from ion repellor grid wires in the detector.  A one-dimensional map of grid-wire opacity for each detector was shifted from detector coordinates into wavelength space and divided from the flux and error vectors.  Exposure time in the location of grid wires was decreased to $\sim70$\%, giving these pixels less weight in the final coaddition.  We also modify the error and local exposure time at the edges of the detector segments to de-weight flux contributions from these regions.  With four different central wavelength settings per grating, any residual instrumental artifacts from grid-wire shadows and detector segment boundaries should have negligible effect on the final spectrum.  

The exposures are aligned with each other and interpolated onto a common wavelength scale.  One exposure in each grating/detector was picked as a wavelength reference, and the remaining exposures were cross-correlated with it.  The wavelength region of cross-correlation for each case was picked to include a strong ISM absorption feature, and shifts were typically on the order of a resolution element ($\sim0.07$~\AA) or less.  The COS wavelength solution has not yet been rigorously characterized, and we see a systematic shift between strong ISM lines and their expected LSR velocities.  The shift is approximately constant across the COS wavelength range, so we apply a uniform $+0.17$~\AA\ shift to the wavelength vectors ($\sim40$ \kms\ at $\sim1300$~\AA) to bring ISM line centroids to the expected $v_{\rm LSR}\approx0$ seen in many ISM absorbers.

%%%%%%%%%%%%%%%%%%%%%%%%%%%%%%%%%%%%%%%%%%%%%%%%%%
%% Figure 2
\begin{figure*}
  \epsscale{1}\plotone{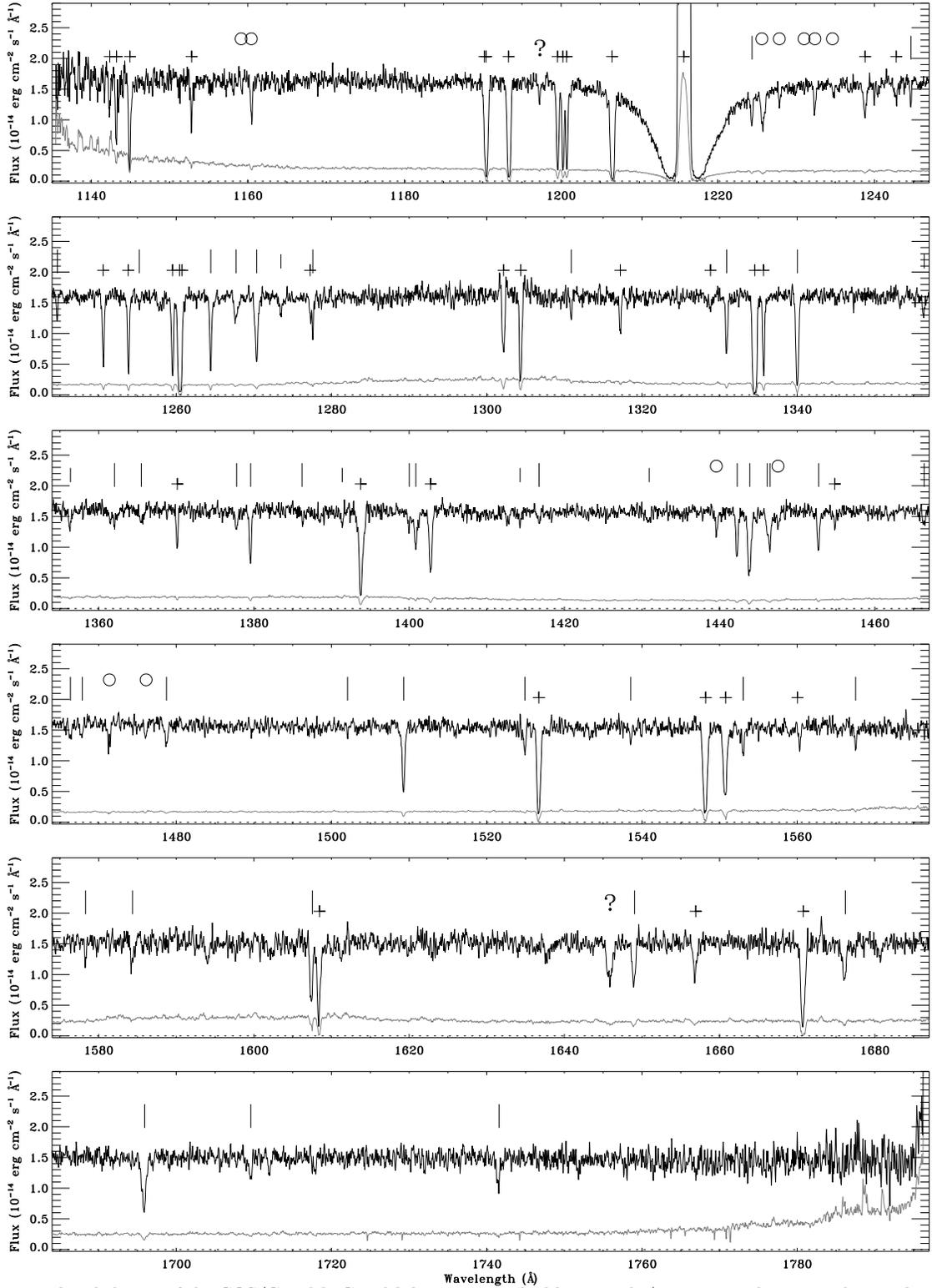} %%%%%%%%%%%%%%%%%%%%%%
  \caption{A more detailed view of the COS/G130M$+$G160M dataset
  smoothed by 7 pixels (approximately one resolution element).  Error
  is shown in gray.  Prominent ISM lines are marked with plus signs.
  IGM \Lya\ absorbers are marked with large vertical ticks.  Smaller
  ticks denote corresponding \Lyb\ detections.  IGM metal absorbers
  are marked with open circles.  The two question marks denote the
  ambiguous features discussed in \S3.  See Table~1 for line
  identifications and measurements.}
\end{figure*}

Next, the aligned exposures were interpolated onto a uniform wavelength grid and coadded.  The flux at each position was taken to be the exposure-weighted mean of flux in each exposure.  Since exposure time was reduced in certain wavelength locations, as noted above, pixels near detector edges and where grid-wire shadows were removed received less weight than those in less suspect locations.  The combined data show $S/N\sim20$ per 7-pixel ($\sim0.07$~\AA) resolution element and are sufficient to detect narrow absorption features down to $W_\lambda\approx15$ m\AA\ at $4\sigma$ significance.  Figure~1 shows the entire combined COS/G130M and COS/G160M spectra.  Figure~2 shows a more detailed view of the spectrum with prominent lines marked.

% FUSE details
In addition to the COS data, we utilize 45 ksec of {\it Far Ultraviolet Spectroscopic Explorer (FUSE)} observations taken 2004 April as part of program E526 (PI: Savage).  While \FUSE\ data alone are insufficient to characterize the \HI\ absorber systems along a sight line, far-UV coverage is invaluable for confirming \Lya\ lines at $z\la0.11$ via \Lyb\ absorption.  Additionally, \OVI\ \lam\lam1032, 1038 and \CIII\ \lam977 absorbers are found only in \FUSE\ data at $z<0.10$ and $z<0.16$, respectively.  Thirty-seven \FUSE\ exposures were retrieved from the archive and processed in the usual manner \citep{Paper1}.  The final \FUSE\ spectrum covers $\lambda=905-1187$~\AA\ with $\rm S/N\approx7-10$ per $\sim20$ \kms\ resolution element.

%%%%%%%%%%%%%%%%%%%%%%%%%%%%%%%%%%%%%%%%%%%%%%%%%%
%% Figure 3
\begin{figure}[b]
\epsscale{.95}\plotone{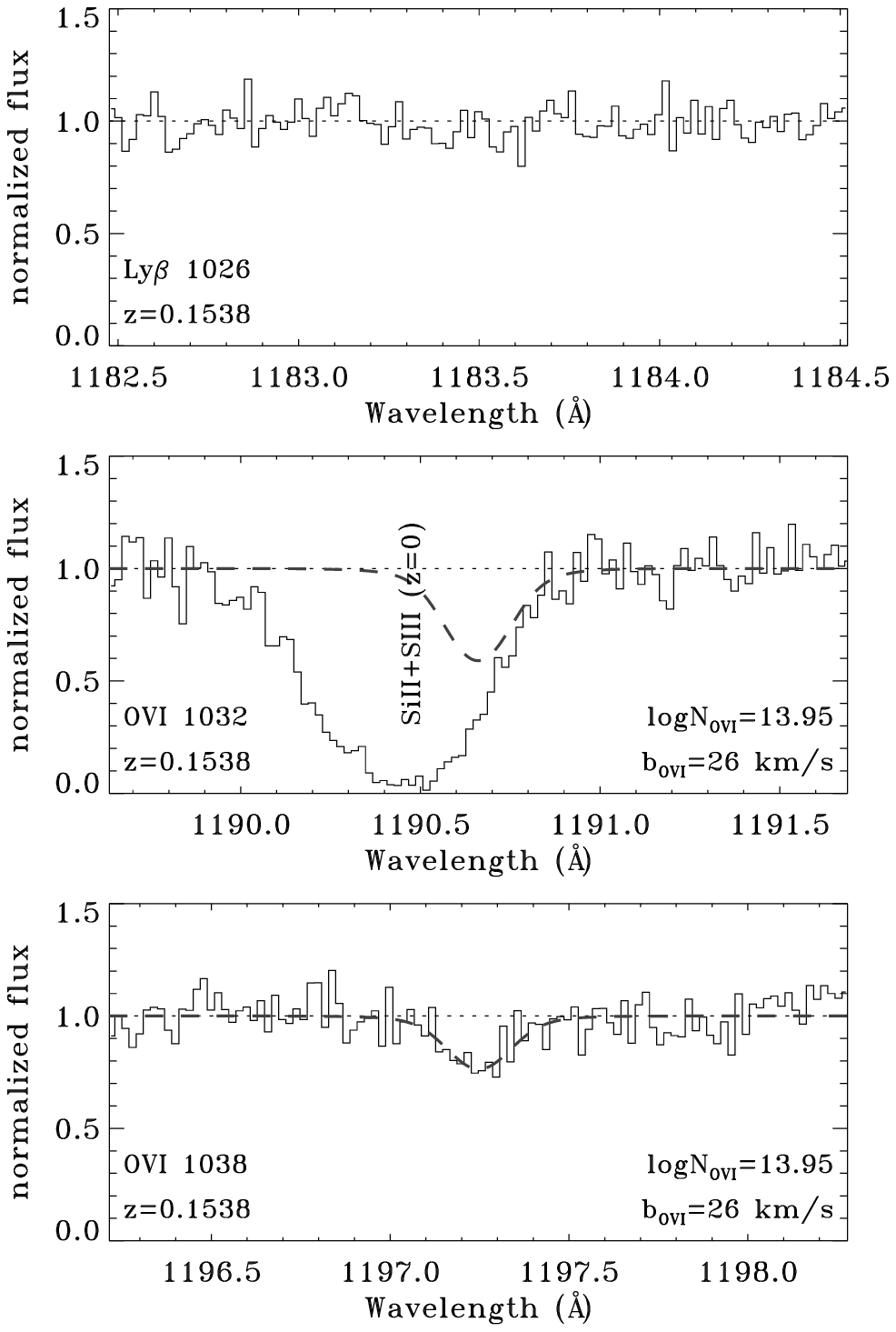} % fig_unknowns.pro %%%%%%%%%%%%%%%%%%%%%%%
  \caption{A weak absorption line at 1197.25~\AA\ ($W_\lambda=50\pm9$
  m\AA) is consistent with O\,VI \lam1038 at $z=0.1538$ (bottom
  panel).  The corresponding O\,VI \lam1032 line is blended with
  Galactic Si\,II$+$S\,III absorption, but consistent with the
  measurement from the O\,VI \lam1038 line (middle panel).  The
  corresponding \Lya\ absorber is blended with Galactic \SiIV\
  \lam1403, but no H\,I is seen in \Lyb\ (top panel) to $4\sigma$
  limit $\log\,N_{\rm HI}<13.81$.}\label{fig:1197id}
\end{figure}

%%%%%%%%%%%%%%%%%%%%%%%%%%%%%%%%%%%%%%%%%%%%%%%%%%%%%%%%%%%%%%%%%%%%%%%%%%%%
\section{Absorption Lines}

An initial analysis of the spectrum reveals a wealth of far-UV absorption features (Figures~1 and~2).  Many of these are clearly Galactic ISM lines typical of most sight lines to Galactic and extragalactic sources.  We label the remainder as redshifted IGM absorbers.  To identify these lines, we follow a procedure similar to that employed in DS08: starting from the long-wavelength end of the spectrum, we interactively mark the strongest absorpton features, tentatively identifying them as \Lya.  The location of the corresponding \Lyb\ absorption is then checked, as are those of prominent metal-ion absorbers (\OVI\ \lam\lam1032, 1038; \CIV\ \lam\lam1548,1550; \SiIII\ \lam1207; \CIII\ \lam977, etc.).  This process is iterative, as we identify weaker and weaker features.  If there is component structure in a line profile that can be unambiguously deconvolved into multiple absorbers, we list these systems separately.  However, most systems are listed as a single absorber, even if they possess rather complex line profiles.  

We note two significant absorption features at 1197.25 \AA\ and 1645.9 \AA\ with highly ambiguous identifications.  The weak feature at 1197.25~\AA\ ($W_{\rm obs}\approx50$ m\AA) cannot be \Lya, nor is it consistent with either a higher-order Lyman line or any obvious metal-ion absorber for any of the known \HI\ systems.  The most plausible identification is that of \OVI\ \lam1038 at $z=0.1538$ (Fig.~\ref{fig:1197id}).  The stronger \lam1032 line of the \OVI\ doublet is blended with Galactic \SiII\ \lam1190.  No \HI\ absorption is seen at this redshift in \Lyb, and a $4\sigma$ upper limit on the column-density can be set at $\log\,N_{\rm HI}<13.81$.  \Lya\ absorption at this redshift is blended with the weaker line of the Galactic \SiIV\ doublet at \lam1403.  However, the Galactic \SiIV\ lines appear in the expected 2:1 ratio, leaving little room for additional blended \Lya\ $z=0.1538$ absorption.  It appears possible that this is a WHIM absorber with high enough temperature and metallicity that no neutral gas is seen \citep[see also][]{Savage10}. 

%%%%%%%%%%%%%%%%%%%%%%%%%%%%%%%%%%%%%%%%%%%%%%%%%%
%% Figure 4
\begin{figure*}[t]
\epsscale{.8}\plotone{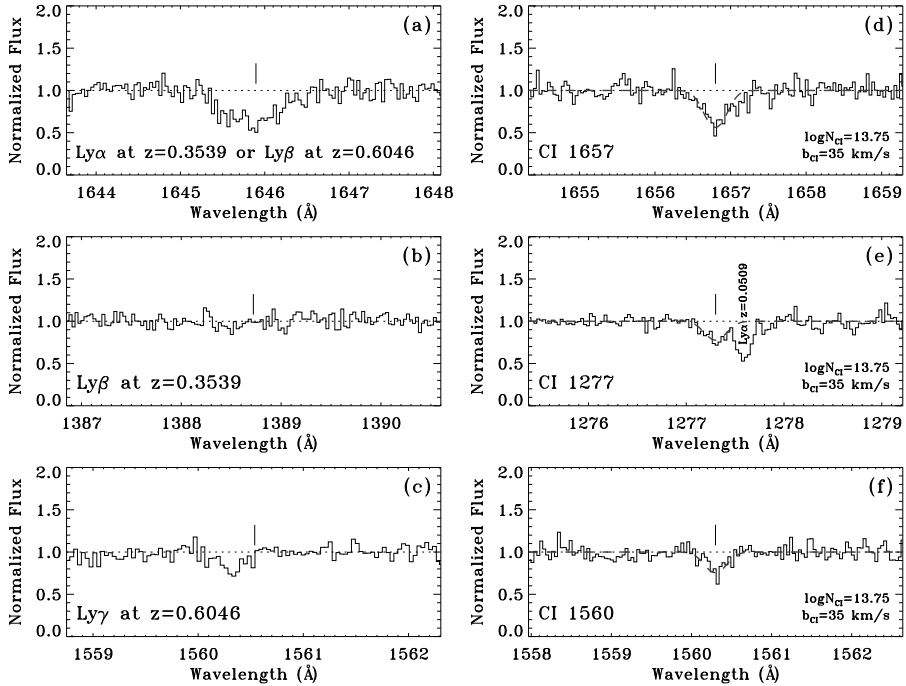} % fig_1645.pro %%%%%%%%%%%%%%%%%%%
  \caption{A broad feature at 1645.9~\AA\ (panel a) is interpreted as
  \Lya\ at $z=0.3539$.  However, the expected \Lyb\ feature (panel b)
  is not seen in the data.  If the 1645.9~\AA\ feature is instead
  interpreted as \Lyb\ at $z=0.6046$, a feature of approximately the
  correct strength is seen at 1560.3~\AA, the location of the expected
  \Lyg\ at $z=0.6046$ (panel c).  However, the 1560.3~\AA\ feature is
  consistent with Galactic C\,I absorption (dashed profile in panels
  d-f).  We conclude that the 1645.9~\AA\ feature is most likely a
  multi-component \Lya\ absorber at $z\approx0.3539$, but the
  individual components are too weak to appear in \Lyb.}
  \label{fig:1645id}
\end{figure*}

The strong absorption line at 1645.9~\AA\ ($W_{\rm obs}\approx357$ m\AA) could be identified as \Lya\ at $z=0.3539$, but the expected \Lyb\ absorber ($W_{\rm obs}\ge50$ m\AA) is not seen (Fig.~\ref{fig:1645id}a,b).  The line is consistent with being \Lyb\ absorption at $z=0.6046$, and an equivalent \Lyg\ feature is seen at 1560.3~\AA\ at approximately the expected strength (Fig.~\ref{fig:1645id}a,c).  However, the latter feature is consistent with Galactic \CI\ absorption lines seen elsewhere in the data (Fig.~\ref{fig:1645id}d-f).  Therefore, we tentatively identify this feature as a multi-component \Lya\ system at $z=0.3539$.  Two or more \Lyb\ features of the required strengths can plausibly be hidden in the noise at the required location.  

Table~1 lists measurements for all detected IGM absorption lines, grouped by redshift, and including the two ambiguous cases above.  In total, we identify 42 IGM absorbers (Table~1), 41 of which are detected in at least \Lya.  Corresponding higher-order Lyman line and/or metal ion absorption is seen in 15 absorbers.  Seven systems show metal absorption.  The observed \Lya\ absorber frequency per unit redshift, $d{\cal N}/dz\approx87\pm15$, down to a limiting equivalent width of 50 m\AA\ ($\sim10^{13}\rm~cm^{-2}$), is similar to that found for the larger DS08 sample to the same limit ($d{\cal N}/dz=95\pm5$).  A more thorough search for broad \Lya\ absorbers with $b>40$ \kms\ will be conducted, following the receipt of additional data on this source scheduled for Cycle~18.  Therefore, we caution the reader that this line list may not be complete for lines with $b>40$ \kms.

%%%%%%%%%%%%%%%%%%%%%%%%%%%%%%%%%%%%%%%%%%%%%%%%%%%%%%%%%%%%%%%%%%%%%%%%%%%%
\section{Results and Discussion}

An additional six orbits of COS integration time planned for Cycle 18 should improve the S/N of the combined dataset by a factor of $\sim2$.  Greatly improved S/N, as well as our evolving understanding of the COS instrumental effects, will enable us to reliably measure low-contrast absorbers such as broad \Lya\ systems and weak metal lines.  We defer a more exhaustive analysis of the sight line until then, but note two key results here.

%%%%%%%%%%%%%%%%%%%%%%%%%%%%%%%%%%%%%%%%%%%%%%%%%%
%% Figure 5
\begin{figure*}[t]
\epsscale{.95}\plotone{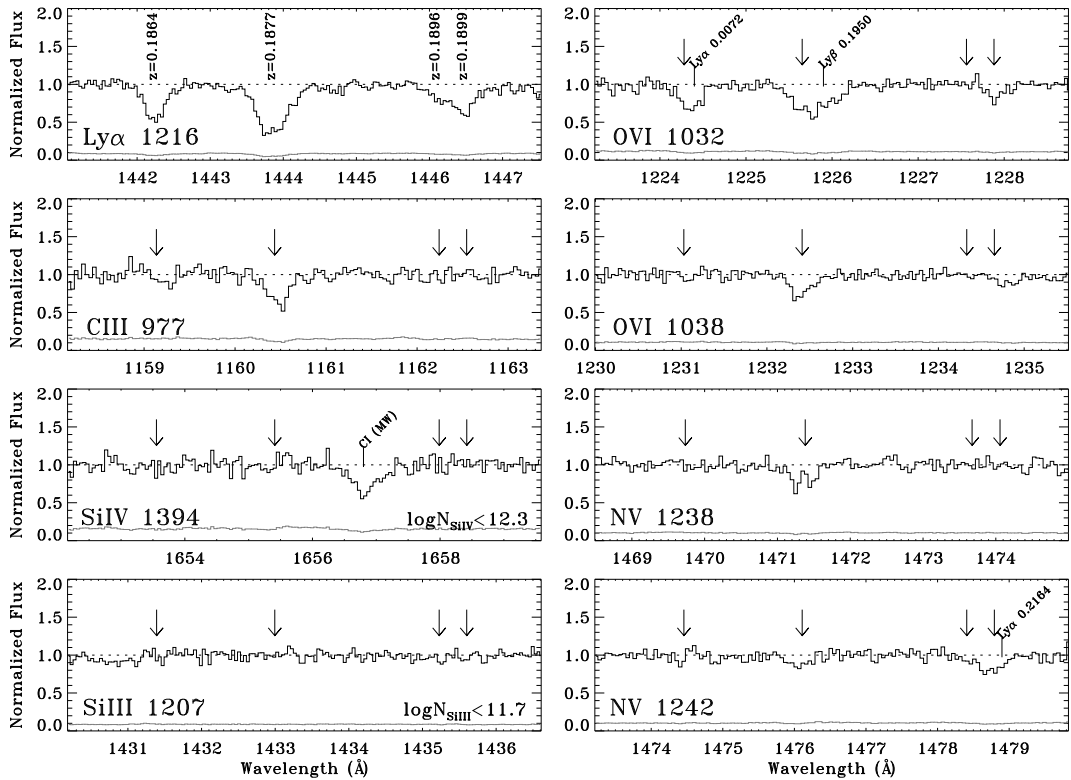} % z01875.pro ~~~~~~~
  \caption{Detail of the absorption complex at $z=0.188$.  Top left
  panel shows three strong \Lya\ absorption systems at $z=0.1864$,
  $z=0.1877$, and $z=0.1897$.  The reddest of these is clearly split
  into two components ($z=0.18958$ and $z=0.18989$).  The strong,
  central absorber also shows clear evidence of multiple structure,
  but is harder to deconvolve unambiguously.  In subsequent panels,
  these components are marked by arrows.  Corresponding metal
  absorption is detected in several of these components in C\,III,
  O\,VI, and N\,V, however there is no measurable absorption in either
  Si\,III or Si\,IV.  Several unrelated lines appear in the profiles
  and are labeled with vertical ticks.  Data is binned by four pixels
  ($\sim$50\% of a resolution element).}\label{fig:z0187}
\end{figure*}

\subsection{Triple Absorber Complex at $z=0.188$}

The most interesting of the previously undiscovered IGM absorption systems is the triplet of metal-rich absorbers at $z=0.18640$, $0.18773$, and $0.18989$ (Figure~\ref{fig:z0187}).  In \Lya, the strong central absorber at $z=0.18773$ is flanked by two weaker components at $z=0.18640$ and $z=0.18989$, or $v=-399$ \kms\ and $v=+648$ \kms, respectively, relative to the system at $z=0.18773$.  

\begin{figure}
  \epsscale{1}\plotone{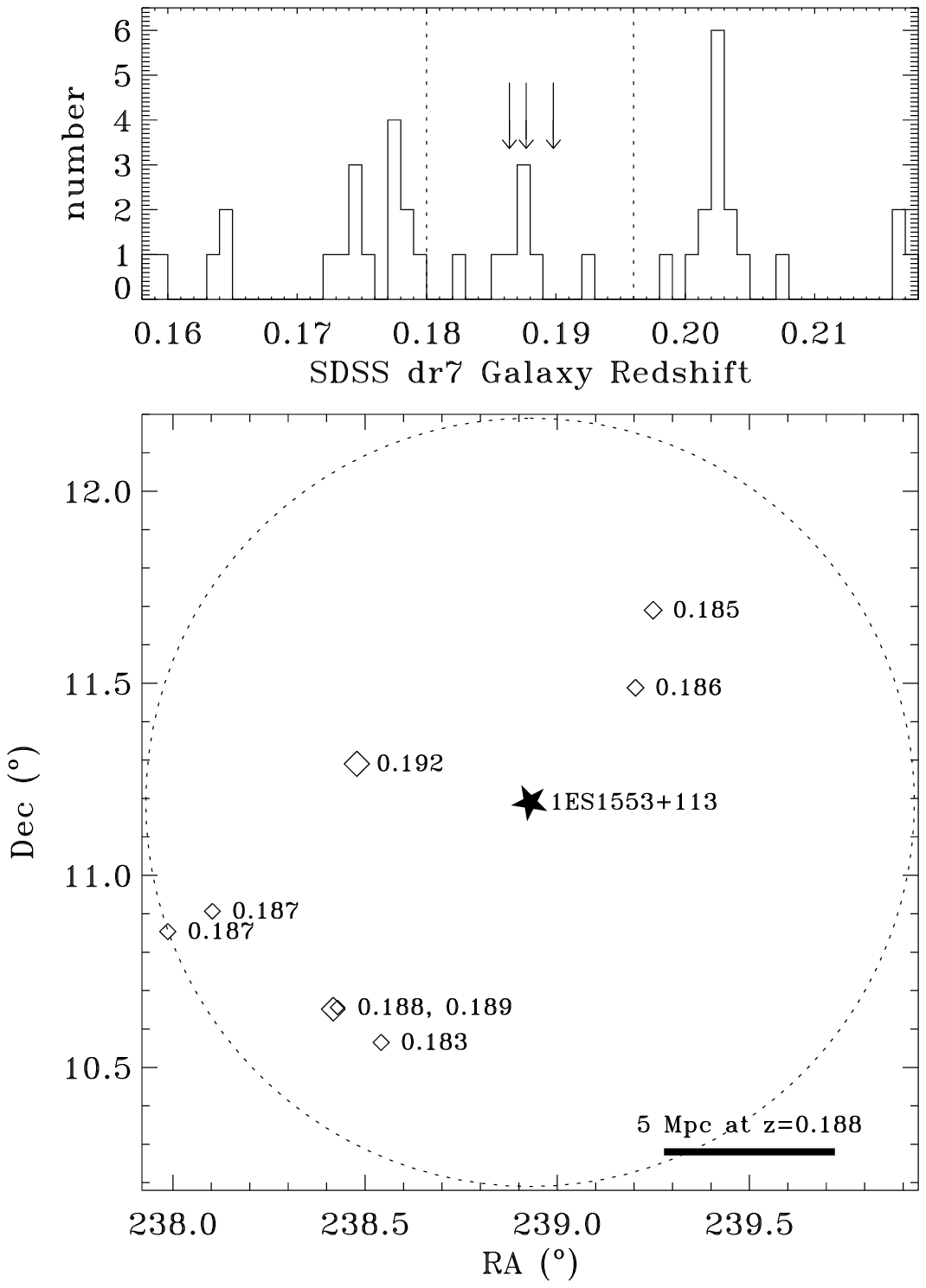} % galsurvey.pro %%%%%%%%%%
  \caption{Correlating the triple absorption system at $z\sim0.188$
  with SDSS galaxies within one-degree of 1ES\,1553$+$113.  The top
  panel shows a the redshift distribution of SDSS galaxies within the
  search radius.  The three \Lya\ absorber redshifts (arrows)
  correspond to a significant peak in the galaxy distribution bounded
  by the vertical dotted lines ($0.180<z<0.196$).  We plot these eight
  galaxies (diamonds) in relation to the 1ES\,1553$+$113 sight line
  (star) in the lower panel.  Spectroscopic redshifts are noted next
  to each diamond, and symbol size denotes galaxy luminosity.  The
  distribution of bright galaxies at approximately the redshift of the
  triple absorption system ($z\sim0.188$) suggests a large-scale
  filament.  Though none of the SDSS galaxies are closer than
  24\arcmin\ ($\sim4.5$ Mpc at $z=0.188$) to the line of sight, the
  SDSS galaxy catalog is not complete to lower-luminosity objects at
  this redshift.  All galaxies in this field have $L\ga3\,L^*$.  A
  deep galaxy redshift survey will likely show fainter galaxies closer
  to the AGN line of sight.}\label{fig:galsurvey}
\end{figure}

The three absorption systems at $z\sim0.188$ span $\sim1000$ \kms\ in comoving velocity space, appropriate for a large-scale filament in the cosmic web.  We searched the SDSS Data Release 7 galaxy redshift catalog \citep{Abazajian09} for galaxies within one degree of the 1ES\,1553$+$113 sight line and plotted them as a function of redshift (Fig.~\ref{fig:galsurvey}, top panel).  While the SDSS is complete only in the brightest galaxies ($L\ga3\,L^*$) at this redshift, a clear concentration appears at $z=0.187\pm0.003$.  None of these galaxies is closer than 24\arcmin\ ($\sim4.5$ Mpc at $z=0.188$) to the line of sight, so it is hard to claim a specific galaxy-absorber relationship (Figure~\ref{fig:galsurvey}, bottom panel).  However, the median redshift of the galaxy sample is $z=0.187$ and the $1\sigma$ deviation ($\sigma_z=0.0027$) is roughly one-third of the redshift search space ($\Delta z=\pm0.008$).  This tight clustering around the absorber redshift, as well as the observed spatial distribution of the brightest galaxies, suggests that the galaxies trace a large-scale filament in the cosmic web and that the absorption in the COS observations arises in the same structure.  Deeper galaxy survey work (Keeney \etal, in preparation) is complete to much lower luminosity ($L\ga0.3\,L^*$) and may show a closer galaxy-absorber relationship.

%%%%%%%%%%%%%%%%%%%%%%%%%%%%%%%%%%%%%%%%%%%%%%%%%%
%% Figure 6

\OVI\ absorption is seen in all three systems ($\log\,N_{\rm OVI}=13.4\pm0.3$, $14.1\pm0.1$, and $13.5\pm0.1$, respectively).  Strong \NV\ absorption is seen in the central component ($\log\,N_{\rm NV}=13.7\pm0.1$).  DS08 measure $\rm \langle N_{\rm NV}/N_{\rm OVI}\rangle=0.24^{+0.22}_{-0.12}$ in eleven \OVI$+$\NV\ low-$z$ IGM absorbers, so the ratio observed at $z=0.18773$ toward 1ES\,1553$+$113 ($\rm N_{\rm NV}/N_{\rm OVI}=0.4$) is high but within the observed range.  The DS08 sample is a biased toward higher \NNV\ values, the observed ratio suggests an elevated N/O abundance in this absorber.

\CIII\ is detected in the central and blue components ($\log\,N_{\rm CIII}=13.25^{+0.06}_{-0.04}$ and $12.64\pm0.15$, respectively), but $4\sigma$ upper limits of $\log\,N_{\rm SiIII}<11.64$ and $\log\,N_{\rm SiIV}<12.29$ can be placed on Si-ion absorption in all three systems.  \SiII\ \lam1260 is tentatively detected at $z=0.1877$ as a pair of weak, narrow components with a total column density $\log\,N_{\rm SiII}\sim12.1$.  However, we do not detect other \SiII\ lines, nor equivalent absorption in \CII\ \lam1334.5 or \CII\ \lam1036.3 ($\log\,N_{\rm CII}\le12.82$) and other singly-ionized species.  

It is likely that the gas in the central $z=0.1877$ \Lya\ system is multi-phase in nature, with a warm-hot ionized medium (WHIM) component traced by \OVI\ and \NV\ and a cooler, photoionized component traced by \HI\ and \CIII.  The ``multiphase ratios'' for these absorbers \citep{Paper1,DS08} are $N_{\rm HI}/N_{\rm OVI}\sim 1.3$, $\sim0.6$, and $\sim0.5$ for the three main components.  Typical values for absorbers with similar \NHI\ are $\sim0.6$, $\sim2.5$, and $\sim0.8$, respectively \citep{Paper1}.  We can use the \Lya\ and low-ionization metal detections and upper limits to constrain metallicity and relative abundances in the photoionized gas.  In particular, \CIII\ and \SiIII\ have similar ionization potentials and are often detected in the same systems.  At solar abundance ratios \citep{Asplund09}, carbon is 8.3 times more abundant than silicon, but \SiIII\ is detectable to much lower column densities than \CIII\ owing to the very strong $f$-value of the 1206.5~\AA\ transition \citep{Shull09}.  Thus, the two ions are often seen together in photoionized IGM absorbers (DS08).

We measure $N_{\rm CIII}/N_{\rm SiIII}>40$ in the $z=0.1877$ absorber, an unusually high lower limit.  DS08 report \CIII\ and \SiIII\ detections in 22 low-$z$ IGM systems with a median distribution of $N_{\rm CIII}/N_{\rm SiIII}=8.5^{+20.4}_{-5.5}$ ($1\sigma$).  In Galactic High Velocity Clouds \citep{Fox06,Shull09} the ratio, $N_{\rm CIII}/N_{\rm SiIII}$, typically ranges from 5--20.  Thus, the abnormally high ratio $\sim 40$ found in these IGM absorbers is well outside the usual range.  Comparing our measurements with a grid of simple CLOUDY photoionization models \citep[detailed in][]{Paper2}, we see that the relative column densities of \CIII\ and \SiIII\ are fairly insensitive to photoionization parameter $U\equiv n_\gamma/n_H$.  Typical model ratios are $(N_{\rm CIII}/N_{\rm SiIII})\sim10$ in the expected range of IGM photoionization parameters ($U\sim10^{-2}$) and are largely insensitive to assumptions about metallicity, photon continuum, and gas density.

The unusually high \CIII/\SiIII\ ratio suggests that the C/Si abundance in this system may have a strongly non-solar abundance pattern.  If \NSiIII\ is typically $\sim$10\% of \NCIII, as seen in other IGM observations and models, $\rm[Si/C]>0.6$, or greater than four times the solar value.  Comparing the observed \NCIII/\NHI\ ratio with the models, we expect $\log\,(N_{\rm CIII}/N_{\rm HI})\la-1.7$ for $Z=0.1\,Z_\sun$.  However, the observed column density ratio is an order of magnitude higher, suggesting that (C/H) is close to solar values in this system.  Without additional low-ionization species detected, we cannot determine whether carbon is overabundant or silicon is underabundant relative to the solar ratio.  The \CIII\ detection at $z=0.1864$ is factor of four weaker than in the main absorber, while the upper limit on \NSiIII\ is the same.  Therefore, this system puts weak constraints on the metallicity and abundances in this absorber.  

Although $\rm(C/Si)=8.3\pm1.2$ in the Sun \citep{Asplund09}, variations in this abundance ratio can occur, depending on the youth of the stellar population and its initial mass function (IMF).  Carbon is produced primarily by helium burning in intermediate-mass stars (red giants, horizontal branch), whereas silicon arises from more advanced $\alpha$-process nucleosynthesis in massive stars.   The usual abundance trends show enhanced (Si/C) and reduced (N/O) in low-metallicity stellar populations \citep{McWilliam97,Cayrel04}.  Theoretical predictions \citep{WoosleyWeaver95} show that [$\alpha$/Fe] increases with increasing progenitor mass (here, $\alpha$ includes O, Mg, Si, S, Ca, Ti).  Thus, a low Si and O abundance compared to C and N suggests an IMF skewed toward low-mass stars.  

Comparing \HI\ or \CIII\ to high ions \OVI\ and \NV\ requires them to be in the same thermal phase for a meaningful analysis.  Hybrid ionization modeling (CIE plus photoionization) of the high ions in this system are reported by \citet{Yao10}, in which the $z=0.1877$ system is used as a test case for a physical parameter-based absorption line modeling exercise.

\subsection{Constraining the Redshift of 1ES\,1553$+$113}

The redshift of 1ES\,1553$+$113 is crucial to determining the intrinsic properties of the source.  Indirect methods of constraining the redshift of 1ES\,1553$+$113 fall into two categories.  First, the ratio of AGN to host galaxy optical luminosity in BL\,Lacs is thought to cover a fairly small range \citep{Wurtz96,Sbarufatti05}.  Various deep ground-based \citep{HutchingsNeff92,Scarpa00} and space-based \citep{Urry00,Carangelo03,Sbarufatti06,Treves07} optical studies have failed to detect a host galaxy beneath the glare of the AGN.  From these non-detections, redshift limits from $z>0.09$ to $z>0.78$ have been set by various groups.  \citet{Treves07} refined this to $z=0.3-0.4$ using a more sophisticated analysis of the same optical data.  However, the validity of the assumption of host/nuclear luminosity relationship has been called into question by \citet{OdowdUrry05}.

A complementary technique uses the observed very high energy (VHE) spectrum ($0.1-10^3$ GeV) to place limits on the redshift of BL\,Lacs.  This method assumes that the VHE spectral energy distribution (SED) of an object will be modified, as TeV photons interact with photons in the ambient extragalactic background and produce $e^+e^-$ pairs \citep[e.g.,][]{YoungerHopkins10,PersicDeAngelis08}.  The longer the pathlength, the steeper the VHE SED becomes.  Uncertainties in the extragalactic IR background and the intrinsic SED of the AGN render this method uncertain, but the redshift of 1ES\,1553$+$113 has variously been constrained to $z<0.74$ \citep{Ahronian06,Albert07} or $z<0.80$ or $z<0.42$ \citep{MazinGoebel07} based on HESS and MAGIC observations.  \citet{Abdo10} use data from the first six months of {\it Fermi} $\gamma$-ray observations in conjunction with observations from radio wavelengths to 1 TeV to model the intrinsic SED of 1ES\,1553$+$113.  Based on these models, they determine a redshift $z=0.75^{+0.04}_{-0.05}$.  The error bars on this estimate appear to be much smaller than justified by this method.

%%%%%%%%%%%%%%%%%%%%%%%%%%%%%%%%%%%%%%%%%%%%%%%%%%
%% Figure 7
\begin{figure*}[t]
  \epsscale{.9}\plotone{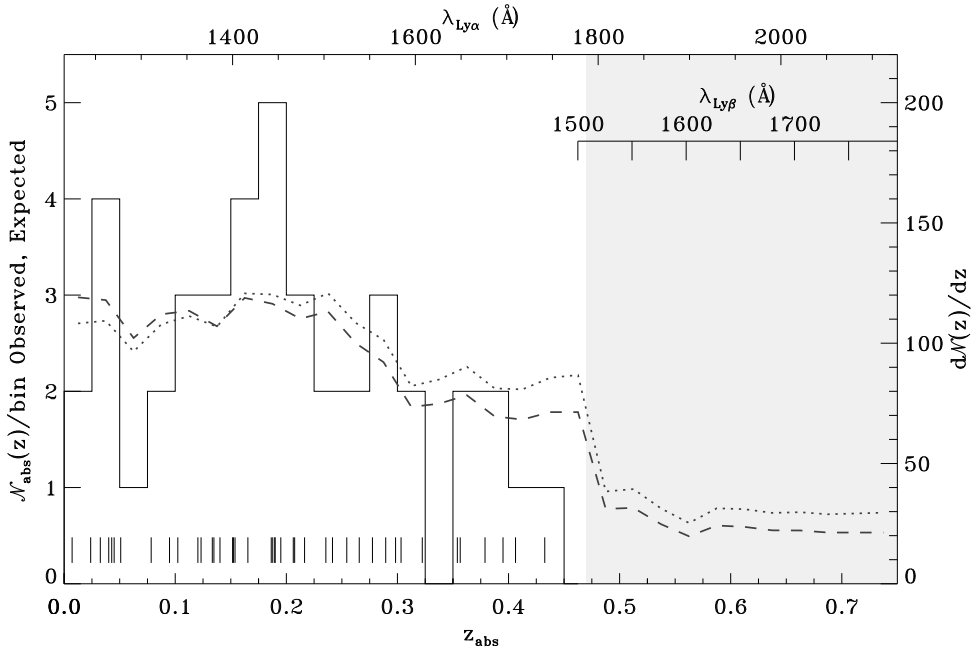} % ~/igm/1es1553/zhist.pro
  \caption{Observed and expected distribution of IGM absorbers along
  the 1ES\,1553$+$113 sight line.  Vertical ticks mark the redshift of
  observed IGM absorbers along the sight line, and the solid line
  shows a histogram of ${\cal N}_{\rm abs}$ per $\Delta z=0.025$
  redshift bin.  We calculate the $10\sigma$ minimum equivalent width
  at each redshift bin based on the observed S/N of the data and plot
  the expected $d{\cal N}/dz$ to that limit (dashed curve) based on
  the large \Lya\ sample of DS08.  If a modest H\,I absorber evolution
  is assumed, $d{\cal N}_{\rm HI}/dz\propto(1+z)^{0.7}$, the expected
  number of IGM systems rises by $\sim20-50$\% at higher redshifts
  (dotted curve).  At $z>0.47$ (shaded region), \Lya\ absorbers can no
  longer be detected in the COS far-UV band and we must rely on much
  less sensitive \Lyb\ detections.  As discussed in the text, no
  $z>0.4$ \Lyb\ absorbers are detected.} \label{fig:zhist}
\end{figure*}

From Figures 1 and 2, it is clear that there are \HI\ systems throughout the redshift range from $z\sim0$ to near the end of the COS spectral coverage ($z=0.47$).  A strong line at 1695~\AA\ is identified as \Lya\ at $z=0.395$ and confirmed by detection of \OVI\ at the same redshift in both lines of the doublet.  This sets a firm lower limit on $z_{\rm em}>0.395$.  Two weaker features at 1709.5~\AA\ and 1741.5~\AA\ appear in the data, which we identify as \Lya\ at $z=0.4063$ and $z=0.4326$, respectively.  Though we do not detect higher-order Lyman or metal-ion lines at these redshifts, the two $z>0.4$ \Lya\ absorbers are weak enough that we do not expect confirmation in other lines.  The continuum of the BL\,Lac object remains smooth across the entire COS band (Figs.~1 and~2), and no intrinsic emission or absorption is seen.  

Thus, we can confidently constrain the emission redshift of 1ES\,1553$+$113 to $z_{\rm em}>0.400$, and it appears likely that it may be as high as $z_{\rm em}=0.433$.  Additional COS observations may detect corresponding \Lyb\ absorption to these two $z>0.4$ absorbers ($W_{\rm Ly\beta}\sim12$ m\AA\ is expected).  The confirmed and unconfirmed direct redshift limits from the HST/COS observations are compatible with both the lower limits set by the non-detection of an optical host galaxy \citep[$z_{\rm em}\ga0.1-0.4$;][]{Urry00,Sbarufatti06,Treves07} and the VHE SED upper limits upper limits \citep[$z_{em}\la0.8$;][]{Ahronian06,Albert07,MazinGoebel07}.  

% higher redshift probe via Lyb, OVI
We now assess the validity of the most recent VHE SED redshift estimate ($z_{\rm em}=0.75^{+0.04}_{-0.05}$) from \citet{Abdo10}.  Our current COS far-UV spectra (G130M and G160M) are only sensitive to \Lya\ absorbers at $z<0.47$.  However, higher redshift absorbers can be detected using the less sensitive higher-order Lyman lines or \OVI\ doublet.  COS far-UV data cover the wavelength range 1135~\AA\ $<\lambda<$1795~\AA\ corresponding to \Lyb\ redshifts $0.11<z<0.75$ and \OVI\ redshifts $0.10<z<0.74$.  \Lyb\ and \OVI\ systems at $z>0.47$ will appear at $\lambda\ga 1508$~\AA.  We find empirically (DS08) that the detection threshhold for absorption lines with no prior ``signposts'' (such as known absorber redshifts) is $\sim10\sigma$.  The COS/G160M data in this region are of sufficiently high quality that we would expect to detect lines of $W_{\rm obs}\ga50$ m\AA\ at a $\sim10\sigma$ significance level.  This corresponds to rest-frame $W_{\rm r}\ga30$ m\AA\ for \Lyb\ absorbers at $z\sim0.5$ or $\log\,N_{\rm HI}\ga13.6$ for \Lyb.  

Figure~\ref{fig:zhist} shows the observed and predicted distribution of IGM absorbers as a function of redshift.  We calculate the expected number of absorbers ${\cal N}_{\rm abs}$ per $\Delta z=0.025$ bin (alternately, \dndz) based on the S/N-determined minimum equivalent width in each bin and the sample of $\sim650$ \HI\ absorbers from DS08.  The dashed curve in the figure represents no \dndz\ evolution with redshift.  The evolution of low-$z$ \HI\ absorbers is somewhat uncertain.  However, if we assume that the \HI\ absorber frequency evolves as $(d{\cal N}/dz)_{\rm HI}\propto(1+z)^\gamma$ \citep{Penton04} and adopt $\gamma\sim0.7$ for a modest evolution between $z=0$ and $z\sim1$, the expected number of \HI\ detections (dotted curve in Figure~\ref{fig:zhist}) rises at higher redshift by $\sim20-50$\%.  The sharp drop in expected detections at $z>0.47$ coincides with the switch from \Lya\ to \Lyb\ as an IGM tracer (shaded area) and the resulting loss of sensitivity discussed above.  Summing the expected number of \Lyb\ absorbers (${\cal N}_{\rm abs,exp}$) which should appear at $\lambda>1508$~\AA\ data from Figure~\ref{fig:zhist} over the range $0.47<z_{\rm abs}<0.75$, we find ${\cal N}_{\rm abs,exp}\sim7$ and ${\cal N}_{\rm abs,exp}\sim9$ in the constant and evolved \HI\ models, respectively.  Thus, we should expect $\sim8$ high-$z$ \Lyb\ absorbers in the 1ES\,1553$+$113 data if $z_{\rm em}\ge0.75$.

Are the predicted high-$z$ \Lyb\ absorbers seen?  As discussed in \S3, the strong feature at 1645.9~\AA\ can be ruled out as \Lyb\ at $z=0.6046$ since the corresponding \Lyg\ line is consistent with Galactic \CI\ (Fig.~\ref{fig:1645id}).  Thirteen other absorption features longward of 1507~\AA\ have been identified as \Lya\ lines, and eight of these identifications are not confirmed with higher-order Lyman or metal-ion lines.  If these eight single-line detections are instead interpreted as potential high-$z$ \Lyb\ systems, six have inconsistent \Lyg\ non-detections and a seventh has \Lyg\ blended with another line.  The only possible high-$z$ \Lyb\ absorber is a marginal line detected at 1584.3~\AA\ and identified as \Lya\ at $z=0.30328$.  If this line is instead identified as \Lyb\ at $z=0.54463$, the corresponding \Lyg\ absorber should be at 1502.2~\AA, nearly concident with a weak feature identified as \Lya\ at $z=0.23559$.  The relative strengths of the two features are consistent with \HI\ absorbers at $z=0.54463$, but both line detections are of relatively low significance.  Additional COS observations may improve the significance of the line detections.  However, we find none of the other predicted high-$z$ \HI\ or \OVI\ absorbers in the data, so the $z_{em}>0.545$ redshift limit from the possible \Lyb\ detection for 1ES\,1553$+$113 is very speculative.  

The additional far-UV observations of 1ES\,1553$+$113 scheduled for HST Cycle 18 will improve the S/N of the current dataset by a factor of $\sim1.7$ and consequently lower the minimum detectable line strength for potential \Lyb\ absorbers at $z>0.47$ by a similar factor.  However, observations with the COS/G185M grating covering the wavelength range $\rm 1800~\AA\la\lambda\la2100~\AA$ ($0.47\la z_{\rm abs}\la 0.73$ in \Lya) would be more efficient at detecting $z>0.47$ IGM absorbers in the sight line.  Despite the relatively lower efficiency of the COS near-UV gratings and detectors compared with their far-UV counterparts, the \Lya\ lines will be $\sim3-7$ times stronger than \Lyb\ and should be easily detected or ruled out with only a few kiloseconds of observations.  Such observations could potentially also measure weak intrinsic broad-line \Lya\ emission from the BL\,Lac object, as has been seen in other BL\,Lac objects at lower redshift (Danforth, Stocke, Winter, \etal, in prep).

We constrain the source redshift of 1ES\,1553$+$113 statistically by truncating the ${\cal N}_{\rm abs, expected}$ model curves in Figure~\ref{fig:zhist} at a range of $0.4<z_{\rm em}<0.75$.  Applying a Kolmogorov-Smirnov test to the different models, we set a $1\sigma$ constraint of $z_{\rm em}\le0.58$ for the non-evolved model (and $z_{\rm em}\la0.49$ for the evolved \HI\ distribution).  An emission redshift of $\sim0.75$ is ruled out for both models at a $90$\% or greater level of confidence.  Thus, we constrain the redshift of 1ES\,1553$+$113 to the range $0.43<z_{\rm em}\la0.58$.

\section{Conclusions}

The BL\,Lac object 1ES\,1553$+$113 is one of the brightest objects in the sky in $\gamma$-rays, as well as being a notable UV and X-ray source.  However, the AGN emission is that of a relativistic jet aligned closely with our line of sight and, like most such objects, has no intrinsic emission or absorption features at any wavelength.  This featureless, power-law continuum is ideal for measuring intervening IGM features that are weak and broad, such as thermally-broadened \Lya\ systems.  However, the lack of intrinsic features makes constraining the redshift of the object difficult.  

We present unprecedented high-quality far-UV HST/COS and \FUSE\ spectra of the BL\,Lac object 1ES\,1553$+$113 at spectral resolution 15-20 \kms.  These data show 42 intervening IGM absorbers, 41 of which are detected in \Lya, and 15 in \Lyb\ and/or metal lines.  The richest absorption system in the line of sight is a trio of \Lya\ absorbers at $z\approx0.188$ covering $\sim1000$ \kms\ of velocity space.  Several metal ions are also detected in these systems, including \OVI, \NV, and \CIII.  However, neither \SiIV\ nor \SiIII\ is detected in any of the systems.  The \CIII/\SiIII\ ratio implies a (C/Si) abundance at least four times the solar value, while a high \NV/\OVI\ value suggests an overabundance of N as well.  A detailed analysis of the physical conditions in this system can be found in \citet{Yao10}.  

% new redshift limit
The redshift of 1ES\,1553$+$113 has never been determined directly, and the only limits placed on it come from indirect means such as the shape of the $\gamma$-ray spectrum and the lack of an AGN host galaxy in deep optical images.  A strong \Lya$+$\OVI\ absorber at $z=0.3951$ gives the first direct lower limit to the redshift of the object.  Two weaker \Lya\ absorbers at $z=0.4063$ and $z=0.4326$ give slightly higher estimates of the redshift, but these weak \Lya\ lines are not confirmed by additional line detections.  

These lower limits are consistent with most previous measurements via optical non-detections of host galaxies and $\gamma$-ray SED constraints.  \citet{Abdo10} derive $z_{\rm em}=0.75^{+0.04}_{-0.05}$ based on the latest {\it Fermi} and TeV $\gamma$-ray SED measurements, considerably higher than our intervening absorber upper limits.  COS far-UV spectra are not sensitive to \Lya\ absorbers at $z>0.47$, but the G160M grating has some sensitivity to intervening \Lyb\ and \OVI\ absorbers out to $z\sim0.75$.  If the \citet{Abdo10} redshift estimate were accurate, we would expect to find $\sim8$ \Lyb\ absorbers at $0.47<z<0.75$.  We find no evidence for any higher redshift absorption systems.  There are only a few absorption features at $\lambda>1500$~\AA\ with ambiguous line identifications that could potentially be \Lyb\ systems at $z>0.47$.  While these systems are individually suggestive, we find nowhere near the number of absorbers predicted statistically.  We conclude that the redshift of 1ES\,1553$+$113 is not much higher than $z\approx0.45$.

% X-ray follow-up
1ES\,1553$+$113 is one of the brightest X-ray sources on the sky and has been suggested as a sight line that could be efficiently probed for WHIM absorption in \OVII.  The combined \OVI\ column density in the three absorbers at $z\sim0.19$ is $\sim2\times10^{14}\rm~cm^{-2}$.  Spectrographs on modern X-ray observatories are sensitive to $\log\,N_{\rm OVII}\ga15.5$.  If the temperature of any of these \OVI\ systems is high enough, sufficiently long Chandra and/or XMM/Newton observations may reveal a \OVII\ counterpart to these \OVI\ absorbers that could constrain the long-sought X-ray WHIM \citep{Bregman07}.  However, at the observed Li-like (\OVI) oxygen column density, $\log\,N_{\rm OVI}\approx14.3$ in the trio of absorbers, the expected column densities of He-like (\OVII) and H-like (\OVIII) oxygen are probably just below the detectability levels of {\it Chandra} and {\it XMM}.  Recent analysis of stacked X-ray absorption data \citep{Yao09} at the known IGM redshifts of \OVI\ absorbers finds no evidence for \OVII\ or \OVIII\ absorbers to a limit $N_{\rm OVII}/N_{\rm OVI}<10$.  Therefore, the $z\approx0.19$ absorbers might have X-ray column densities $\log\,N_{\rm OVII}\leq15.3$, just below the limits of current X-ray observatories. 

% further COS data
Finally, these observations showcase the powerful new tool available to astronomers for probing the low-redshift IGM.  COS is $10-20$ times more sensitive in the far-UV to point sources than previous instruments on HST.  An additional six orbits of COS observations are planned for 2010, which should improve the S/N of the combined dataset by a factor of $\sim\sqrt{3}$.  Improving the data quality will help confirm or refute some of the tentative line identifications from this paper and will undoubtedly uncover additional weak absorbers.  We will place further constraints on [C/Si] and [N/O] in the $z=0.188$ system, identify new broad, shallow \Lya\ absorbers, and investigate possible high-$z$ \Lyb\ systems with our new Cycle 18 observations.

\medskip
\medskip
% Acknowledgements
It is our pleasure to acknowledge the many thousands of people who made the HST Servicing Mission 4 the huge success that it was.  We furthermore thank Steve Penton, St\'ephane B\'eland, and the other members of the COS ERO and GTO teams for their work on initial data calibration and verification.  C. D. wishes to acknowledge a fruitful discussion with members of the KIPAC consortium.  This work was supported by NASA grants NNX08AC146 and NAS5-98043 to the University of Colorado at Boulder.

\vspace{3cm}

%%%%%%%%%%%%%%%%%%%%%%%%%%%%%%%%%%%%%%%%%%%%%%%%%%%%%%%
%% Table 1
\LongTables 
\begin{deluxetable*}{lccccl}
\tabletypesize{\scriptsize}
\tablecolumns{6} 
\tablewidth{0pt} 
\tablecaption{Intergalactic Line Detections}
\tablehead{\colhead{$z_{\rm abs}$}   &
           \colhead{Line}    &
           \colhead{$W_{\rm r}$ (m\AA)}    &
           \colhead{$b\rm~(km\,s^{-1})$}     &
           \colhead{$\log\,N$} &
           \colhead{notes}    }
	\startdata
0.00717&Ly$\alpha$&$108\pm12$&$ 35\pm 4$&$13.37\pm0.04         $&blend with O\,VI $z=0.187$ \\
0.02386&Ly$\alpha$&$ 54\pm 8$&$ 19\pm 4$&$13.07\pm0.05         $& \\
0.03254&Ly$\alpha$&$ 24\pm 8$&$ 13\pm 7$&$12.70\pm0.10         $& weak \\
0.04012&Ly$\alpha$&$215\pm 9$&$ 27\pm 1$&$13.86\pm0.02         $& \\
       &Ly$\beta$\tablenotemark{a} &$ 87\pm46$&$ 11\pm 6$&$14.39\pm0.22         $&ambiguous ID \\
       &          &          &$ 17^{+ 21}_{ -5}$&$14.25^{+0.78}_{-0.47}$   & H\,I COG solution\tablenotemark{b} \\
0.04281&Ly$\alpha$&$135\pm14$&$ 73\pm 7$&$13.44\pm0.04         $& broad \\
0.04498&Ly$\alpha$&$275\pm10$&$ 47\pm 2$&$13.87\pm0.01         $& asymmetric \\
       &Ly$\beta$\tablenotemark{a} &$ 90\pm 8$&$ 45\pm 2$&$14.19^{+0.09}_{-0.09}$& \\
       &    &  &$ 24   ^{+  3}_{ -3}$&$14.20\pm0.07         $   & H\,I COG solution\tablenotemark{b} \\
0.05094&Ly$\alpha$&$ 84\pm 9$&$ 17\pm 3$&$13.32\pm0.04         $& \\
0.07832&Ly$\alpha$&$ 58\pm10$&$ 24\pm 6$&$13.09\pm0.06         $& weak \\
0.09481&Ly$\alpha$&$176\pm 9$&$ 30\pm 2$&$13.67\pm0.02         $& \\
       &Ly$\beta$\tablenotemark{a} &$ 34\pm 8$&$ 26\pm11$&$13.80^{+0.17}_{-0.19}$& weak, ambiguous ID \\
       &    &  &$ 28^{+\infty}_{-10}$&$13.70^{+0.13}_{-0.14}$   & H\,I COG solution\tablenotemark{b} \\
0.10230&Ly$\alpha$&$360\pm 4$&$ 52\pm 1$&$14.12^{+0.03}_{-0.03}$&BLA? \\
       &Ly$\beta$\tablenotemark{a} &$ 81\pm10$&$ 36\pm12$&$14.20^{+0.12}_{-0.11}$& \\
       &    &  &$ 42   ^{+ 12}_{ -6}$&$14.10\pm0.08         $   & H\,I COG solution\tablenotemark{b} \\
0.12040&Ly$\alpha$&$ 32\pm 7$&$ 21\pm 9$&$12.85^{+0.07}_{-0.14}$& weak \\
0.12325&Ly$\alpha$&$ 77\pm17$&$ 81\pm16$&$13.18\pm0.07         $& weak, broad \\
       &Si\,III 1206&$ 16\pm 8$&$  9\pm 8$&$11.93\pm0.14       $& tentative \\
0.13334&Ly$\alpha$&$ 70\pm13$&$ 46\pm 8$&$13.15\pm0.06         $& \\
0.13483&Ly$\alpha$&$156\pm 9$&$ 30\pm 2$&$13.60\pm0.02         $& \\
0.14028&Ly$\alpha$&$ 28\pm 8$&$ 18\pm 8$&$12.76\pm0.10         $& weak \\
0.15164&Ly$\alpha$&$ 51\pm 9$&$ 28\pm 6$&$13.02\pm0.06         $& \\
0.15234&Ly$\alpha$&$173\pm13$&$ 66\pm 5$&$13.56\pm0.03         $& \\
0.15380&O\,VI 1038  &$ 50\pm9 $&$ 26\pm 6$&$13.95\pm0.06         $& tentative, see \S3, Fig. \ref{fig:1197id} \\
0.16540&Ly$\alpha$&$ 35\pm 8$&$ 40\pm11$&$12.83\pm0.09         $& weak \\
0.18640&Ly$\alpha$&$139\pm 7$&$ 32\pm 2$&$13.52\pm0.02         $& \\
       &Ly$\gamma$&$ 10\pm 7$&$ \sim5  $&$13.84\pm0.56         $& weak \\
       &    &  &$ 17^{+\infty}_{ -6}$&$13.67^{+0.27}_{-0.24}$   & poorly-constrained H\,I COG\tablenotemark{b} \\
       &O\,VI 1038  &$ 10\pm 4$&$ \sim5  $&$13.40\pm0.24         $& low significance \\
       &C\,III 977  &$ 26\pm11$&$ 22\pm14$&$12.64\pm0.15         $& noisy \\
0.18773&Ly$\alpha$&$293\pm 8$&$ 50\pm 1$&$13.89\pm0.01         $& double \\
       &Ly$\gamma$&$ 23\pm32$&$ 15\pm12$&$14.02\pm0.27         $& marginal \\
       &    &  &$ 36^{+\infty}_{-14}$&$13.99^{+0.44}_{-0.18}$   & poorly-constrained H\,I COG\tablenotemark{b} \\
       &O\,VI 1032  &$108\pm16$&$ 41\pm 6$&$14.04^{+0.08}_{-0.06}$& complicated structure \\
       &O\,VI 1038  &$ 68\pm 2$&$ 35\pm 1$&$14.11^{+0.04}_{-0.04}$& \\
       &N\,V 1238   &$ 74\pm 7$&$ 36\pm 4$&$13.62^{+0.07}_{-0.04}$& two components \\
       &N\,V 1242   &$ 50\pm11$&$ 36\pm 9$&$13.71\pm0.08         $& single component \\
       &C\,III 977  &$ 89\pm 6$&$ 36\pm 5$&$13.25^{+0.06}_{-0.04}$& two components \\
       &Si\,II 1260 &$ \sim18 $&$ \sim5  $&$12.1\pm0.2$           & two components, tentative\\
0.18958&Ly$\alpha$&$ 85\pm17$&$ 49\pm13$&$13.24\pm0.11         $& \\
0.18989&Ly$\alpha$&$ 74\pm 8$&$ 28\pm 4$&$13.20\pm0.05         $& \\
       &Ly$\gamma$&$ 21\pm10$&$  7\pm10$&$14.03\pm0.15         $& weak \\
       &O\,VI 1032  &$ 38\pm 8$&$ 21\pm 6$&$13.53\pm0.07         $& \\
       &O\,VI 1038  &$ 16\pm 5$&$ 13\pm 9$&$13.48^{+0.14}_{-0.19}$& \\
0.19503&Ly$\alpha$&$124\pm 7$&$ 32\pm 2$&$13.45\pm0.02         $& \\
0.20621&Ly$\alpha$&$ 58\pm12$&$ 44\pm10$&$13.06\pm0.07         $& weak \\
0.20747&Ly$\alpha$&$ 53\pm12$&$ 44\pm11$&$13.02\pm0.08         $& weak \\
0.21640&Ly$\alpha$&$ 87\pm14$&$ 46\pm 7$&$13.25\pm0.05         $& \\
0.23559&Ly$\alpha$&$ 20\pm 7$&$ 12\pm 8$&$12.62\pm0.12         $& weak \\
0.24155&Ly$\alpha$&$242\pm10$&$ 36\pm 2$&$13.84\pm0.02         $& strong, multi-component? \\
       &Ly$\beta$ &$ 34\pm 5$&$ 26\pm 7$&$13.73^{+0.07}_{-0.10}$& weak \\
0.25439&Ly$\alpha$&$ 68\pm 2$&$ 31\pm 2$&$13.17^{+0.04}_{-0.05}$& \\
0.26560&Ly$\alpha$&$ 46\pm10$&$ 26\pm 7$&$12.97\pm0.08         $& weak \\
0.27753&Ly$\alpha$&$ 77\pm10$&$ 26\pm 4$&$13.22\pm0.05         $& \\
0.28944&Ly$\alpha$&$ 56\pm10$&$ 24\pm 6$&$13.07\pm0.07         $& \\
0.29830&Ly$\alpha$&$ 56\pm12$&$ 19\pm 5$&$13.09\pm0.07         $& \\
0.30328&Ly$\alpha$&$ 70\pm 7$&$ 34\pm 6$&$13.20^{+0.06}_{-0.08}$& weak, noisy \\
0.32235&Ly$\alpha$&$190\pm 4$&$ 34\pm 4$&$13.73\pm0.04         $& \\
       &Ly$\beta$ &$ 55\pm 9$&$ 38\pm 7$&$13.92\pm0.06         $& \\
       &    &  &$ 18   ^{+  4}_{ -2}$&$13.98^{+0.10}_{-0.11}$   & H\,I COG solution\tablenotemark{b} \\
0.35390&Ly$\alpha$&$235\pm21$&$ \sim79 $&$13.73\pm0.03$         & tentative, see \S3, Fig.~\ref{fig:1645id} \\
0.35650&Ly$\alpha$&$170\pm14$&$ 43\pm 4$&$13.60\pm0.03         $& \\
       &Ly$\beta$ &$ 31\pm 9$&$ 28\pm 9$&$13.66\pm0.10         $& \\
       &    &  &$ 28^{+\infty}_{-12}$&$13.66^{+0.15}_{-0.16}$   & poorly-constrained H\,I COG\tablenotemark{b}\\
0.37884&Ly$\alpha$&$136\pm19$&$ 47\pm 6$&$13.47\pm0.06         $& \\
       &Ly$\beta$ &$ 23\pm 5$&$ 15\pm 5$&$13.55\pm0.08         $& \\
       &    &  &$ 27^{+\infty}_{-15}$&$13.54\pm0.12         $   &poorly-constrained H\,I COG\tablenotemark{b} \\
0.39505&Ly$\alpha$&$278\pm18$&$ 57\pm 4$&$13.84\pm0.02         $&multiple components \\
       &Ly$\beta$ &$ 38\pm 1$&$ \sim54 $&$13.76^{+0.07}_{-0.05}$&broad, multi-component \\
       &O\,VI 1032  &$ 57\pm 6$&$ 31\pm 4$&$13.71\pm0.04         $& \\
       &O\,VI 1038  &$ 27\pm 4$&$ 23\pm 5$&$13.69^{+0.09}_{-0.08}$& \\
0.40630&Ly$\alpha$&$ 59\pm 5$&$ 29\pm 2$&$13.13^{+0.11}_{-0.06}$& weak \\
0.43261&Ly$\alpha$&$ 89\pm 6$&$ 32\pm 2$&$13.31^{+0.07}_{-0.05}$& noisy\\
	\enddata
\tablenotetext{a}{Measurement based on \FUSE\ data.}
\tablenotetext{b}{Curve of growth $b_{\rm HI}, N_{\rm HI}$ solution based on multiple Lyman lines.}

\end{deluxetable*}

\end{document}